\documentclass[prb,twocolumn,a4paper,floatfix,superscriptaddress,unsortedaddress,showpacs,showkeys]{revtex4-1}

\usepackage{dcolumn,amsmath,xspace}
\usepackage{graphicx}
\usepackage{subfigure}
\usepackage{amssymb}
\usepackage{epstopdf}

\newcommand{\degC}{\ensuremath{~^{\circ}\text{C }}}

\begin{document}
\title{Silicon intercalation into the graphene-SiC interface}

\author{F. Wang}
\author{K. Shepperd}
\author{J. Hicks}
\author{M.S. Nevius}
\author{H. Tinkey}
\affiliation{The Georgia Institute of Technology, Atlanta, Georgia 30332-0430, USA}
\author{A. Tejeda}
\affiliation{Institut Jean Lamour, CNRS - Univ. de Nancy - UPV-Metz, 54506 Vandoeuvre les Nancy, France}
\affiliation{Synchrotron SOLEIL, L'Orme des Merisiers, Saint-Aubin, 91192 Gif sur Yvette, France}
\author{A. Taleb-Ibrahimi}
\affiliation{UR1 CNRS/Synchrotron SOLEIL, Saint-Aubin, 91192 Gif sur Yvette, France}
\author{F. Bertran}
\author{P. Le F\`{e}vre}
\affiliation{Synchrotron SOLEIL, L'Orme des Merisiers, Saint-Aubin, 91192 Gif sur Yvette, France}
\author{D.B. Torrance}
\author{P. First}
\author{W.A. de Heer}
\affiliation{The Georgia Institute of Technology, Atlanta, Georgia 30332-0430, USA}
\author{A.A. Zakharov}
\affiliation{Maxlab, Lund University, S-22100, Lund, Sweden}

\author{E.H. Conrad}
\affiliation{The Georgia Institute of Technology, Atlanta, Georgia 30332-0430, USA}

\begin{abstract}
In this work we use LEEM, XPEEM and XPS to study how the excess Si at the graphene-vacuum interface reorders itself at high temperatures. We show that silicon deposited at room temperature onto multilayer graphene films grown on the SiC$(000\bar{1})$ rapidly diffuses to the graphene-SiC interface when heated to temperatures above 1020\!\degC. In a sequence of depositions, we have been able to intercalate $\sim\!6$ ML of Si into the graphene-SiC interface.
\end{abstract}
\vspace*{4ex}

\pacs{73.22.Pr, 61.48.Gh, 79.60.-i}
\keywords{Graphene, Graphite, SiC, Silicon carbide, Graphite thin film}
\maketitle
\newpage

\section{Introduction\label{S:Intro}}

The interface between epitaxial graphene and bulk SiC plays a dominant role in both the growth and transport properties of graphene on SiC.\cite{de Heer_SSC_07,Speck_MSF_10,Hass_JPhyCM_08} For reasons not understood, graphene growth on the $(000\bar{1})$ (C-face) is much faster than growth on the $(0001)$ (Si-face).\cite{Hass_JPhyCM_08}  Nonetheless, it is widely recognized that the diffusion of Si through the two different SiC-graphene interfaces is responsible for these different graphene growth kinetics.\cite{Hass_JPhyCM_08,Emtsev_NatMat_09,WaltPNAS} In addition to playing an important role in growth, the chemical nature of the interface has a dramatic effect on electron transport in monolayer graphene.  Just like in growth, transport on the Si-face is different than on the C-face.  Mobilities are nearly an order of magnitude higher on C-face graphene compared to Si-face graphene even though the structural quality of the films are apparently identical.\cite{de Heer_SSC_07}

The ability to understand and control the SiC-graphene interface will be a necessary step on the road to graphene electronics.  Indeed work has already begun in this direction.  Previous work on the Si-face of SiC has shown that Ge\cite{Kubler_PRB_05,Emtsev_PRB_11}  and interfacial metals\cite{Gierz_NatM_08} can influence the charge transfer from the SiC to the graphene. Similarly, $\text{H}_2$ absorption has been shown to both neutralize the charge transfer and cause the buffer layer to become isolated from the SiC interface.\cite{Riedl_PRL_09} These intercalants can cause substantial changes to the transport properties of the graphene.  On the Si-face, $H_2$ absorption at the interface can increase electron mobilities by 40\%\cite{Speck_MSF_10} (even higher mobilities are achieved if the samples are gated).\cite{Jobst_H2paper}

In this work we present data showing that large amounts of Si can be intercalated into the graphene-SiC interface by annealing Si deposited at the graphene-vacuum interface.  Intercalation proceeds through heterogeneous sites\cite{Hicks_Review} in the graphene film beginning at $\sim\!970$\!\degC and becomes very rapid by 1020\!\degC. This process offers potentially new avenues to modify the graphene-SiC interface structure for new device geometries.

\section{Experimental\label{S:Exp}}

The substrate used in these studies was an n-doped ($n\!=\!2\times\!10^{18}\text{cm}^{-2}$) 6H-SiC. The sample was grown in a closed RF induction furnace using the Controlled Silicon Sublimation (CSS) method at 1550\!\degC for 20 min.\cite{WaltPNAS}  The sample was transported in air before introduction into the UHV analysis chamber.  Prior to LEEM/XPEEM measurements the graphene films were thermally annealed at 500\!\degC in UHV. LEEM/PEEM measurements were
performed at the I311-PEEM beam line at the MAXlab synchrotron radiation laboratory.

ARPES and XPS measurements were made at the Cassiop\'{e}e beamline at the SOLEIL synchrotron in Gif sur Yvette. The high resolution Cassiop\'{e}e beamline is equipped with a modified Petersen PGM monochromator with a resolution $E/\Delta E \simeq 70000$ at 100 eV and 25000 for lower energies. The detector is a $\pm 15^{\circ}$ acceptance Scienta R4000 detector with resolution $\Delta E\!<\!1$meV and $\Delta k\!\sim\!0.01\text{\AA}^{-1}$ at $\hbar \omega\!=\!36$ eV. The total measured instrument resolution is $\Delta E\!<\!12$meV.

The Si was deposited using direct current heating of a Si wafer at 1150\!\degC.  The Si deposition rate was calibrated in three ways to check for consistency: (i) using the Si vapor pressure and the deposition geometry,  (ii) using the measured Si 2p intensity for multiple depositions assuming a layer by layer growth and (iii)  measuring the attenuation of the graphene-C 1s peak intensity before and after Si deposition. All three ways gave a deposition rate of $0.7\!\pm\!.05$ ML/10 min.

\section{Results\label{S:Results}}

\begin{figure}
\includegraphics[width=7.0cm,clip]{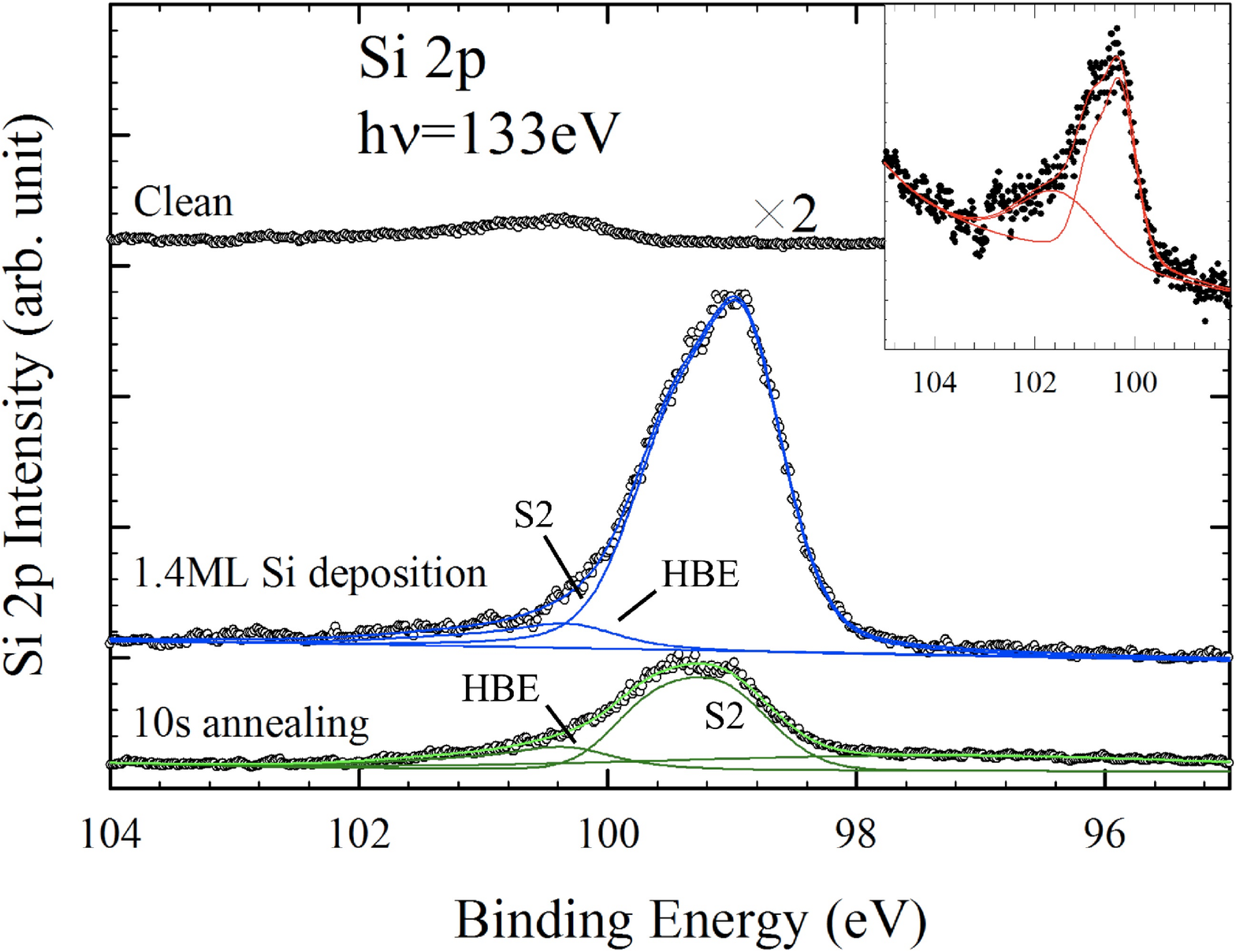}
\caption{Si 2p spectra and fits from the 2 ML graphene region. Spectra are shown for clean, after 1.4 ML of Si was deposited, and after the Si was annealed at 1020\!\degC for 10sec. The insert shows an expanded view of the graphitized spectra before Si deposition. The photon energy was $h\nu=133$eV. Solid lines are the fit peaks and background.} \label{F:SiXPS}
\end{figure}

We have performed a series of Si deposition experiments on top of a 2 ML graphene region of the graphene film using XPEEM.  The 2 ML region was identified using Hibino et al's method of LEEM reflectivity oscillations.\cite{Hibino_PRB_08} The insert in Fig.~\ref{F:SiXPS} shows the Si 2p core level spectra from the as-grown sample.  We have fit the spectra with a spin orbit split Si 2p peak (S2) and a second peak at higher binding energy (HBE) we associate with a SiC shifted surface state similar to the one observed for the clean SiC $(3\!\times\! 3)$ surface.\cite{Sieber_PRB_03} The HBE peak forms after the first anneal and remains after further deposit and anneal cycles.

When 1.4 ML of Si is evaporated at room temperature onto the 2 ML graphene, the Si 2p peak is larger and shifted 1.4eV to lower Binding Energy (BE).  Annealing to 970\!\degC does not significantly affect the spectrum.  However, after a 10 sec anneal to 1020\!\degC, the Si 2p intensity shifts 0.22eV to higher BE and decreases in intensity (see Fig.~\ref{F:SiXPS}). There are four possible explanations for this apparent loss of Si: (i) evaporation of $\sim1$ ML of Si from the surface, (ii) diffusion of a similar amount of Si to heterogeneous defect sites where 3D Si can nucleate (these sites cannot be underneath the photon beam), (iii) Intercalation of Si into the graphene, or (iv) diffusion of Si to the graphene-SiC interface. In cases (i) and (ii) the intensity decrease is simply Si leaving the probe area, while in cases (iii) and (iv) the Si 2p intensity decrease is caused by the electron mean free path that attenuates the photoelectron signal.  As we will now show, Si intercalation into the graphene-SiC interface accounts for most of the lost intensity.

\begin{figure}
\includegraphics[width=7.0cm,clip]{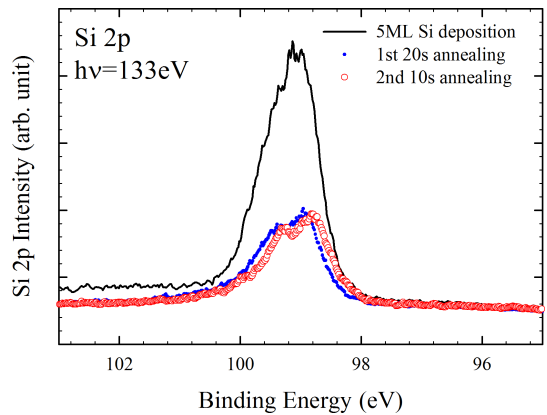}
\caption{Si 2p intensity after 5 ML of Si has deposited (solid line), after the film has been annealed for 20 sec at 1020\!\degC (blue dots), and after an additional 10sec anneal at 1020\!\degC (red open circles). The small shift of peak following the second anneal may results from work function change.} \label{F:60min}
\end{figure}

Significant Si evaporation can be immediately ruled out because the Si evaporation rate at 1020\!\!\degC can account for a loss of no more than $\sim 0.3$ ML.  More importantly, once the intensity has dropped after the first anneal, additional heating at the same temperature has no further effect on the Si 2p intensity (see Fig.~\ref{F:60min}).  This means that after the initial anneal, the Si distribution remains stable.  This stability also rules out substantial diffusion to heterogeneous sites. Some Si does initially diffuses to heterogeneous sites,\cite{Hicks_Review} but the fact that further annealing at the same temperature does not change the Si 2p intensity indicates that this type of material flow quickly becomes negligable.

We performed three similar Si deposit and anneal cycles.  After each anneal, the Si 2p peak drops from its initial deposition value and remains stable with additional annealing to 1020\!\degC.  In addition, after each deposition and anneal cycle, the Si 2p intensity is larger than in the previous cycle. In other words the post-annealed stable Si concentration is building up within the $5\mu$m area probed by the XPEEM.  This buildup is either as Si intercalation between graphene sheets or Si accumulating at the graphene-SiC interface.

We point out that none of the Si is intercalated between graphene sheets. This can be demonstrated by using $\mu$-LEED and taking advantage of the unique stacking of C-face graphene. C-face graphene consists of sheets that are commensurately rotated with non-Bernal angles (i.e not $60^\circ$).\cite{Hass_PRL_08,Sprinkle_JphysD_10}  This stacking leads to graphene super cells that can be seen in STM.\cite{Hass_PRL_08,Miller_Science_09,Miller_NP_10}  Because there are many rotational pairs in a typical LEED beam diameter and because the amplitude of the supercell vertical displacements are small ($\ll\!0.05$nm),\cite{Rutter_PRB_07} typical LEED patterns from the super cells are averaged over multiple weak patters to produce a diffuse background.

\begin{figure}
\includegraphics[width=8.0cm,clip]{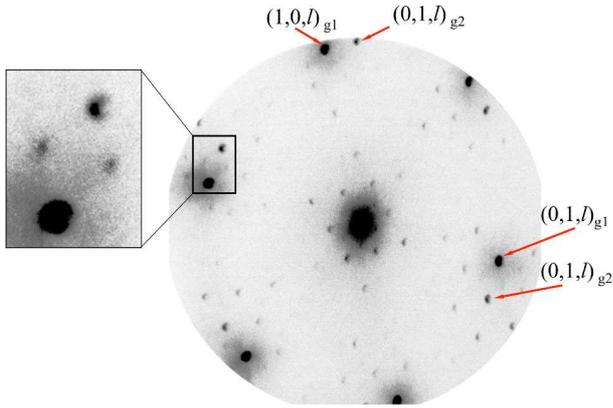}
\caption{$\mu-$LEED image at 96eV from 6H C-face graphene. The LEED image is from the same 2 ML region where the Si 2p data in Figs.~\ref{F:SiXPS} and \ref{F:60min} were taken. The principle diffraction rods from two rotated graphene sheets, are marked.  The insert shows a blow up of the superstructure unit cell.  The cell corresponds to a $(\sqrt{57}\!\times\!\sqrt{57})_G\text{R} 6.59^\circ$ reconstruction.} \label{F:LEED}
\end{figure}

Using $\mu$-LEED, we have selected a sub micron region of the sample to observe the supercell structure before and after Si annealing. The $\mu-$LEED image from the 2-layer region before Si deposition is shown in Fig.~\ref{F:LEED}.  The image shows a pattern that consists of the diffraction pattern from two vertically stacked but rotated graphene layers, $g1$ and $g2$, plus a superstructure pattern. The two graphene sheets are commensurately rotated by $13.17^{\circ}$. This commensurate angle gives rise to the supercell that can be used to index all the superlattice spots (see insert in Fig.~\ref{F:LEED}). The supercell in this region corresponds to $(\sqrt{57}\!\times\!\sqrt{57})_G\text{R} 6.59^\circ$ in graphene units. The superstructure is not due to adsorbates since the structure persists up to at least 1100\!\degC.

Even after annealing 7 ML of Si, there is no effect on the superstructure pattern. If Si intercalated between graphene sheets, the large change in the interlayer spacing caused by inserting Si between graphene would destroy the small corrugation superlattice. After ruling out the other sources of the lost Si 2p intensity, we are left with the conclusion that after annealing,  the majority of the deposited Si is incorporated between the SiC interface and the first graphene layer.

\begin{figure}[hbtp]
\includegraphics[width=9.0cm,angle=90,clip]{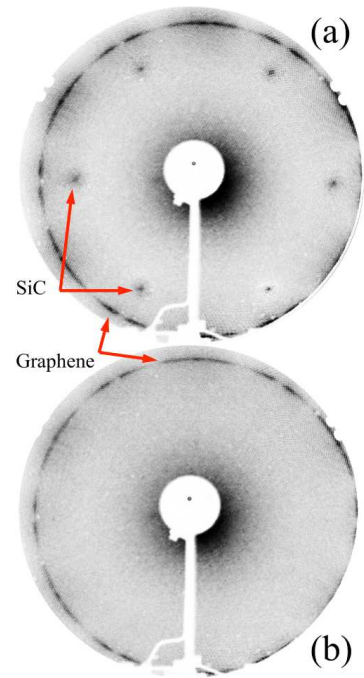}
\caption{LEED patterns from; (a) a masked part of the graphitized surface where Si was not deposited, and (b) from the same sample but at a region where 7 ML of Si was deposited and annealed.  Note that the weak, but visible, SiC $1\!\times\! 1$  pattern in (a) is not visible in (b). Both images were made at 72eV.} \label{F:LEED_comp}
\end{figure}

The most direct evidence that significant Si intercalates into the SiC-graphene interface comes from LEED and XPS measurements.  Figure \ref{F:LEED_comp} shows the LEED pattern from two parts of the same sample after more than 7 ML of Si was deposited and subsequently annealed at 1020\!\degC.  The image in Fig.~\ref{F:LEED_comp}(a) is from the covered part of the sample were no Si was deposited. This is a typical LEED pattern from multilayer graphitized SiC.  It shows both the graphene ring and, because the sample is thin, the diffraction pattern from the $(1\!\times\! 1)$ SiC surface.  Figure \ref{F:LEED_comp}(b) shows a pattern from the part of the sample exposed to Si.  This LEED pattern only shows the graphene ring. Had the Si evaporated or remained on the graphene surface, the diffraction pattern would be the same as in Fig.~\ref{F:LEED_comp}(a) or uniformly attenuated.  Instead only the SiC spots have disappeared.  This demonstrates that a significant portion of the deposited Si has diffused to the SiC-graphene interface where it attenuates the diffraction pattern from the SiC, leaving the graphene pattern unchanged.

\begin{figure}[hbtp]
\includegraphics[width=7.5cm,clip]{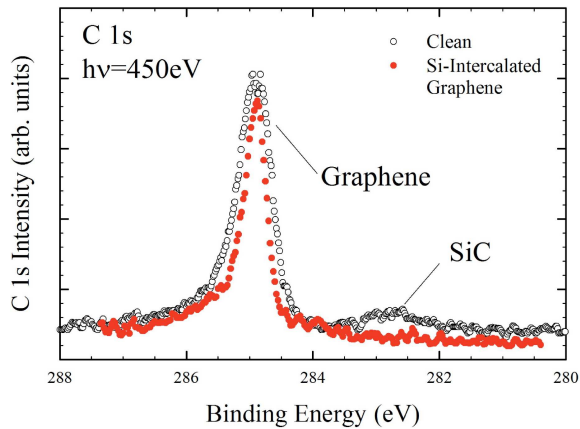}
\caption{C 1s spectra before Si intercalation (open circles) and after intercalation (closed circles).} \label{F:XPS_after}
\end{figure}

A similar effect is seen in the C1s spectra before and after Si intercalation.  Figure \ref{F:XPS_after} shows that the C 1s signal from the SiC carbon disappears from the spectra after intercalation.  The interface Si has attenuated the SiC carbon signal relative to the graphene carbon C 1s photoelectron intensity.

\begin{figure}[hbtp]
 \includegraphics[width=7.5cm,clip,angle=90]{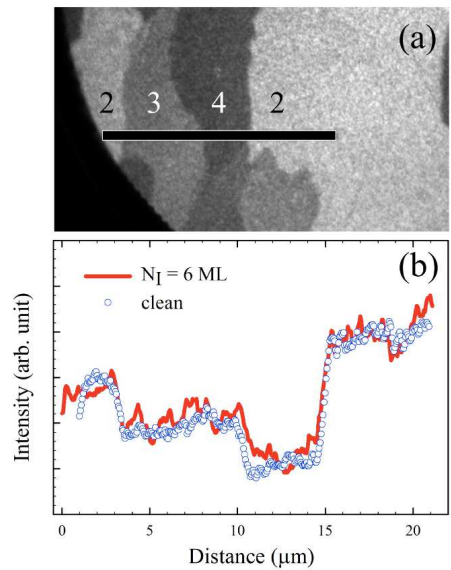}
 \caption{(a) Clean sample XPEEM image using the SiC-Si 2p peak at a BE of 25.8eV with photon energy $h\nu=133$eV. The labels indicate the number of graphene layers determined by LEEM reflectivity.  (b) Intensity profile across the vertical bar in (a) from the clean sample (blue open circles) and after $\sim$7 ML of Si has been deposited and $\sim$6.3 ML intercalated into the interface (solid red line).}
  \label{F:step}
\end{figure}

There is no evidence that the Si converts any significant amount of graphene to SiC at the 1020\!\degC annealing temperature. Figure \ref{F:step}(a) show an XPEEM map using the Si 2p core level intensity.  Because of photoelectron attenuation, image contrast is proportional to graphene thickness (lighter areas are thinner).  Figure \ref{F:step}(b) shows a cut through a 2-,3-,4-, and 2-layer graphene region of the sample before and after Si deposit and annealing.  The plot shows that after Si has moved to the SiC-graphene interface, neither the graphene boundaries nor thicknesses are changing, i.e, there is no measurable conversion of graphene to SiC.

\section{Discusion \label{S:Discus}}

\begin{figure}
\includegraphics[width=7.0cm,clip]{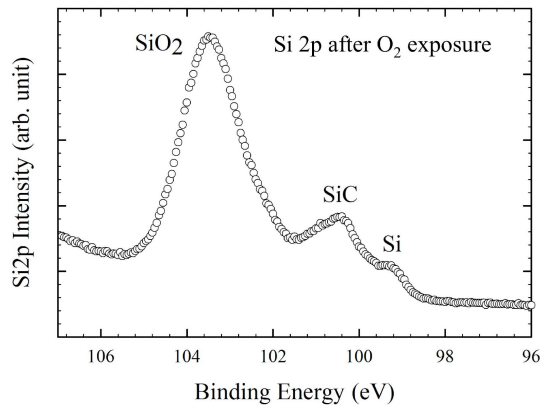}
\caption{The Si 2p core level spectra after the 6.3 ML Si-intercalated sample was exposed to air at room temperature and subsequently heated to 1200\!\degC in vacuum.} \label{F:Si_O2}
\end{figure}

In order to quantitatively determine how much of the deposited Si diffuses to the interface we have performed a detailed analysis of the XPEEM and LEEM data from three deposition and anneal cycles.  We use a model that assumes that most of the Si is uniformly distributed both at the surface and at the SiC interface. LEEM mirror mode images, which are sensitive to height variations on the surface, show no significant height variation after annealing.  Since Si 3D islands on scales larger than the 10nm resolution of the microscope would be visible as surface height fluctuation, we can rule out large scale Si-island formation.  We also allow a fraction of the Si to bind to C sites on the last graphene layer.  We know that a stable Si complex is at the surface because it oxidizes when exposed to air (see Fig.~\ref{F:Si_O2}) and can only be removed by heating to 1450\!\degC. This explains the HBE peak in Fig.~\ref{F:SiXPS}.

\begin{figure}
\includegraphics[width=6.0cm,clip]{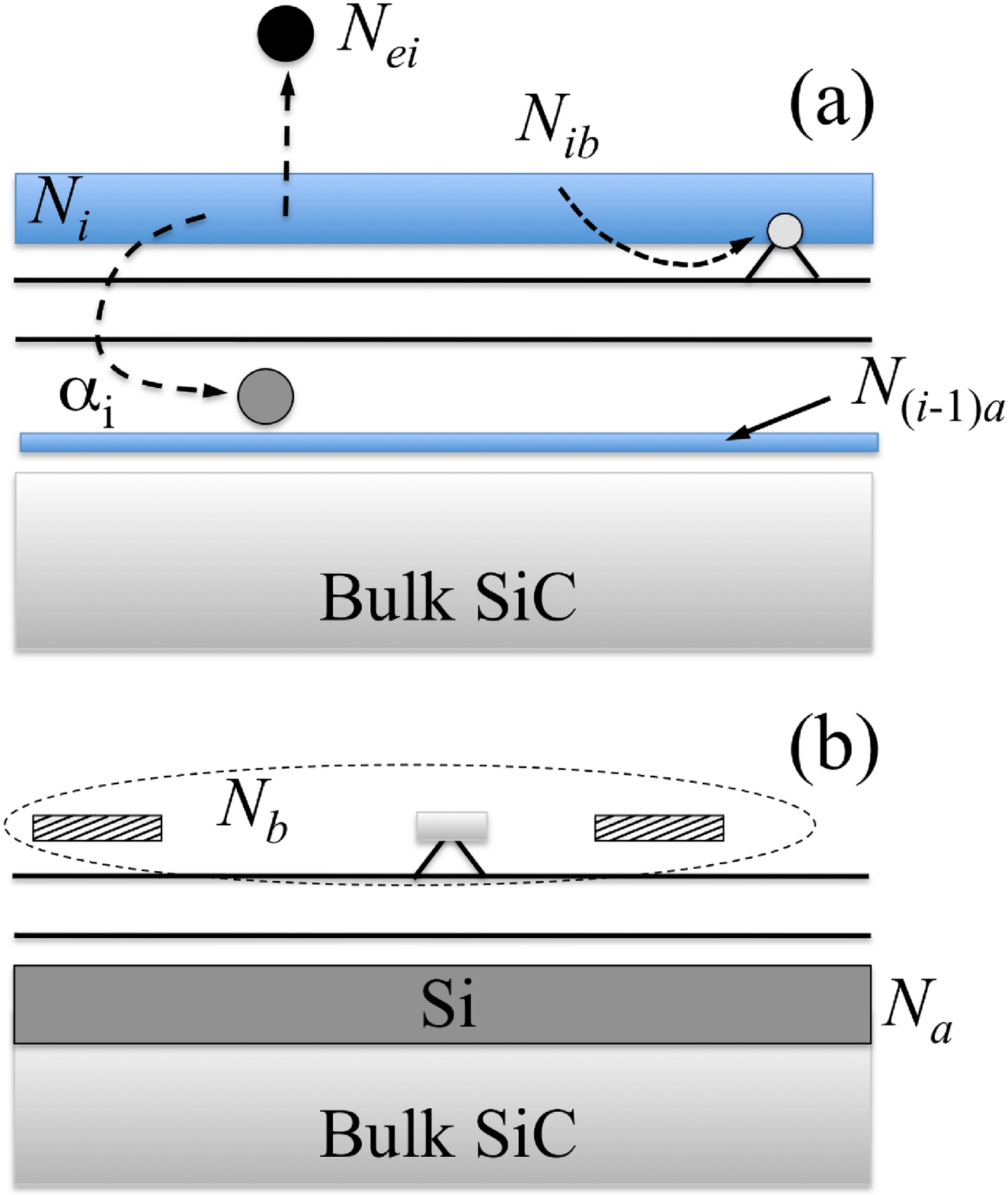}
\caption{(a) Model of how an $N_i$ monolayer Si film is redistributed after annealing.  Some of the surface Si has formed a carbide bonded to the top graphene layer.  (b) Schematic of the surface after 7 ML have been deposited and annealed.  Note that the surface Si and surface carbide are treated as part of the total surface coverage.} \label{F:Model}
\end{figure}

The model used to analyze our experimental results is shown schematically in Fig.~\ref{F:Model}.  In each cycle $N_i$ ($i$=1,2, and 3) monolayers of Si where deposited ($N_1=0.7$ ML, $N_2=1.4$ ML and $N_3=5$ ML). After each anneal, $N_{ei}$ layers are evaporated, $N_{ia}=\alpha_i N_i$ layers have moved to the SiC interface and the remaining $N_{ib}\!=\!(1-\alpha)N_i-N_{ei}$ are bound at the graphene-vacuum interface.  As discussed above, the model does not include intercalation of Si between graphene layers.  The total Si intercalated into the SiC interface is $N_a=\sum_{i}N_{ia}$. The measured Si 2p photo-electron current after annealing, $I'^{(i)} _{Si}$, has contributions from: (i) the Si at the graphene-SiC interface that is attenuated by both the Si within the layer and the graphene above it, and (ii) the total Si at the graphene-vacuum boundary.  The intensity before annealing, $I^{(i)} _{Si}$, depends essentially on the amount of Si deposited because the surface Si and the graphene effectively attenuate any signal from the Si at the SiC interface. This is a reasonable assumption because the electron mean free paths in graphene and Si ($\lambda_g$  and $\lambda_{Si}$, respectively) are small at the kinetic energies used in these experiments (see below).

Assuming that the Si 2p cross section is the same regardless of how it is bound, we can write the relative intensity change after each anneal, $I'^{(i)} _{Si}/I^{(i)} _{Si}$, as;

\begin{widetext}
\begin{subequations}\label{Eq:model}
\begin{equation}
\frac{I'^{(1)} _{Si}}{I^{(1)} _{Si}}=1-\alpha_1+\alpha_1 e^{-2d_g/\lambda_g}[1-(1-\alpha_1)N_{1}(1-e^{-d_{Si}/\lambda_{Si}})]
\label{Eq:model_1}
\end{equation}
\begin{equation}
\frac{I'^{(2)} _{Si}}{I^{(2)} _{Si}}=\frac{N_{2}(1-\alpha_2)+N_{1b}-N_{2e}+(N_{1a}+N_{2}\alpha_2)e^{-2d_g/\lambda_g}{[1-(N_{2}(1-\alpha_2)+N_{1b}-N_{2e})](1-e^{-d_{Si}/\lambda_{Si}})}}{{(1+[N_{2}+N_{1b}-1]e^{-d_{Si}/\lambda_{Si}})}}
\label{Eq:model_2}
\end{equation}
\begin{equation}
\frac{I'^{(3)} _{Si}}{I^{(3)} _{Si}}=\frac{N_{1b}+N_{2b}+(1-\alpha_3)N_{3}-N_{3e}+e^{-2d_g/\lambda_g}/(1-e^{-d_{Si}/\lambda_{Si}})}{1/(1-e^{-d_{Si}/\lambda_{Si}})}
\label{Eq:model_3}
\end{equation}
\end{subequations}
\end{widetext}
where $d_{Si}$ and  $d_g$ are the bulk Si and graphene interlayer spacing, respectively.   In Eq.~\ref{Eq:model_1}  we assume that the evaporated Si in the first step, $N_{1e}$, is zero because of the formation of a partial monolayer stable carbide as discussed above. We also assume that in the third deposition [Eq.~\ref{Eq:model_3}], where the Si interface layer has become very thick (i.e, $N_ad_{Si}\gg\lambda_{Si}$), that we can treat the interface Si as an infinite film.

The mean free path in Si at the kinetic energies used in these experiments (27.7eV) is 3.3\AA.\cite{Himpsel_PRB_88}  The mean free path in graphene at the same low kinetic energy used in this experiment is less well known.  We have estimated $\lambda_g$ by comparing the Si 2p intensity from a 7 ML Si film on graphene to the the Si 2p from SiC attenuated by two graphene layers.  This is done by assuming the Si film has a uniform bulk Si density and the SiC is a bulk terminated surface.  Assuming the photo-electron production cross-section is the same in both layers, we estimate $\lambda_g=2.8\pm0.1${\AA}.  For comparison the mean free path in SiC is 3.2{\AA}.\cite{Sieber_PRB_03}  We have used both values in the coverage calculations.  The variance in the interface Si coverage introduced by using these two values for $\lambda_g$ is $0.1$ ML.

The bulk Si evaporation rate was used as an upper limit estimate for $N_{ei}$ for both the second and third anneal cycles.  This gives  $N_{e1}=N_{e2}=0.3$ ML.  This is small compared to the total deposited film and therefore does not significantly affect the total intercalation value.  It does however influence how much Si remains on the surface.

Using the experimental integrated Si 2p intensities ratios, we can solve Eq.~\ref{Eq:model}(a)-(c) for the three cycles.   The results are summarized in Table \ref{T:Fits}.  A total of 6.3 ML of Si was intercalated into the graphene-SiC interface.  This leaves 0.8 ML that has either evaporated, diffused to defect sites, or remains on the surface.  If we use the total estimated evaporation rate of 0.3 ML, then an upper limit of 0.8-0.6=0.2 ML of Si remains unaccounted for (presumable by diffusion to heterogeneous sites).  A second estimate of the amount of Si remaining on the surface can be made by using the attenuation of the graphene C 1s core level peak before and after intercalation.  This estimate gives 0.5 ML of Si on the graphene film.  This is an upper estimate since part of the remaining Si can (and does) diffuse to heterogeneous sites.

\begin{table}
\caption{\label{T:Fits} Fit parameters for each cycle.  All coverages are in monolayers.}
\begin{ruledtabular}
\begin{tabular}{cccccc}
Cycle  & Deposited Si & $\alpha_i$(\%)& Intercalated &$N_e\footnotemark[1]$ &Surface\\
\hline
1        & 0.7 & 69  &0.5   & 0      &  0.2 \\
2        & 1.4  & 71 & 1.0  &  0.3  &  0.1 \\
3        & 5.0  & 96 & 4.8  &   0.3 &   -0.1\\
Total & 7.1  & 89 & 6.3 &    0.6 &   0.2-0.5\\
\end{tabular}
\end{ruledtabular}
\footnotetext[1]{Estimated from bulk evaporation rate.}
\end{table}

\section{Conclusion\label{S:Conclude}}

We have shown that Si rapidly diffuses from the graphene-vacuum surface to the graphene-SiC interface at 1020\!\degC.  Up to 6.3 ML of Si were intercalated into the graphene-SiC interface in this study.  The reason the Si intercalates is not known.  However, at 1020\!\degC, the SiC phase diagram predicts that any excess Si will stay segregated.  Since the excess Si has the choice of wetting either the graphene at the vacuum boundary or wetting both the graphene and the carbon rich SiC interface, the lower energy choice is intercalation between the SiC and graphene. From an equilibrium view point, the excess Si should bond with the excess carbon (in the form of graphene) to form a new stable SiC layer.  The fact that it does not, indicates that either the surface phase diagram allows both an excess Si and C phase or there is an appreciable kinetic barrier to SiC formation.

A partial Si layer ($\sim\!0.2-0.5$ ML) forms at the vacuum interface during the first deposition and anneal cycle and remains tightly bound to the top layer of graphene during subsequent depositions and anneal cycles. This layer forms an oxide when exposed to air that can only be removed by heating above 1400\!\degC.  There is evidence that above 1400\!\degC, Si can be removed from the interface, but more work needs to be done before this transfer of Si between the vacuum and the SiC interface can be understood.

\begin{acknowledgments}
This research was supported by the W.M. Keck Foundation,  and the NSF under Grant No. DMR-0820382 and DMR-1005880.  We also wish to acknowledge the SOLEIL synchrotron radiation facilities and the Cassiop\'{e}e beamline. J. Hicks also wishes top acknowledge support from the NSF Graduate Research Fellowship Program.
\end{acknowledgments}

%%%%%%%%%%%%%%%%%%%%%%%%%%%%%%%%%%%%%%%%%%%%%%%%%%%%%%%%%%%%%%%%%%%%%%%%%%%%%%%
%%%%%%%%%%%%%%%%%%%%%%%%%%%%%%%%%%%%%%%%%%%%%%%%%%%%%%%%%%%%%%%%%%%%%%%%%%%%%%%

%\newpage

%Attenuation references

%\cite{Seah_SurfIA_79}
%$\Lambda = 3.4$\AA

%\cite{Cumpson_SurfIA_97}
%$\Lambda = 2.5$\AA

%\cite{Barrett_PRB_05}
%$\Lambda = 1.3$\AA

\end{document}